\documentstyle[12pt]{article}
\input epsf
\begin{document}
\title{\bf   Surface Waves and Forced Oscillations
 in QHE Planar Samples  }

\author{{\bf  Alejandro Cabo\thanks{Permanent Address:
Grupo de F\'{\i}sica Te\'orica, Instituto de Cibern\'etica,
 Matem\'atica y F\'{\i}sica,
Calle E, No. 309, Esq. a 15, Vedado, La Habana 4, Cuba}} \\
{\small\it International Centre for Theoretical Physics,}\\
{\small\it P.O.Box. 586,34100, Trieste, Italy},\\
\, \\
{\bf Alejandro Correa,  Ram\'on Guevara}\\
{\small\it Grupo de F\'{\i}sica Te\'orica, 
Instituto de Cibern\'etica, Matem\'atica y F\'{\i}sica,}\\
{\small\it Calle E No.309 Esq. a 15, Vedado,
    La Habana 4, Cuba} ,\\
and \\
{\bf Carlos Rodriguez} \\
{\small \it Departamento de F\'{\i}sica Te\'orica, Facultad de F\'{\i}sica,
 Universidad de La Habana}\\
 {\small\it  Calle San Lazaro, Vedado, La Habana, Cuba}}
\vspace{-3cm}
\maketitle

\begin{abstract}
\noindent
Dispersion relations and polarizations for  surface waves
in infinite planar samples in the QHE regime are explicitly determined
in the small wavevector limit in which the dielectric tensor can
be considered local. The wavelength and frequency regions of
applicability of the results extends to the infrared region for
typical experimental conditions. Then, standard samples with millimetric
sizes seem to be  able to support such excitations. Forced oscillations
are also determined which should be  generated in the 2DEG
by external electromagnetic sources. They show an almost frequency
independent wavevelength which decreases with the magnetic field.
A qualitative model based in these solutions is also presented
to describe
a recently found new class of resonances appearing near the edge of a 2DEG
in the QHE regime.

\end{abstract}

\newpage
\baselineskip=25pt
\section{Introduction}

In spite of  intense research activity  to
determine precisely  the laws for the behavior of charge and
current distributions in samples undergoing the QHE, there remains
 many interesting questions to be answered, and
 experimental results constantly add  new aspects to the
 subject. See for example ~\cite{kohmo}-\cite{ashoo}.

During early activity in the field,
~\cite{mac},\cite{riess},\cite{thou}
 static effective equations for the
electric and magnetic fields were obtained . Already, at that time,
there were predictions that the fields would be
 localized on the boundaries.
Recently experiments have observed this phenomenom in small
mesoscopic samples ~\cite{fontein}. Also in \cite{riess}
 In addition, it was pointed out that the magnetic field should
 contribute to the charge density. This follows
 simply from  charge conservation. These kinds of density
distributions are  essential ingredients of the Chern-Simons approach
proposed early in Refs. \cite{wid}, \cite{ishi}. At that time,
 however, the relevance  of the 3D character of the electromagnetic
 field was not appreciated. Because of this , the merits
of the CS approach were not fully realized. To clarify the way
 in which the CS action is relevant for
the QHE it was argued in ~\cite{cabo} that discussions
 in ~\cite{mac}-\cite{thou}
  could also be considered as being described by  the CS action.
  Simply, this action describes in the Euler equations
   for an electromagnetic field, the
response of a medium showing linear  Hall conductivity.
In addition it was argued that the purely 2D Chern-Simons discussion, with
its predictions of  Meissner effect
   and a massive photon, could be realized in semiconductor superlattices.
The case of time dependent fields was also considered in that  work.
More recently, the importance  of the Chern-Simons action for the
explanation of the electromagnetic response,  also
emerged from the side  of the anyonic and composite fermion approaches
for the QHE ~\cite{girvin},~\cite{kivel},~\cite{frad},~\cite{wen}. Using
the microscopic composite fermion approach effective
electromagnetic equations were derived in \cite{kohmo}, that
 resembled the previous  results of ~\cite{mac}-\cite{thou}.
 This discussion in [1]
supported the conclusions of the former approach ~\cite{mac},~\cite{thou},
~\cite{cabo} by again
stressing the role CS (Hall) currents.
These currents while concentrated near the
border are  different from the purely edge currents that flow very much
closer to it. This fact does not invalidate
the Buttiker picture because both  currents are closely linked by
 gauge invariance ~\cite{wen}.
   
In the present work, we  investigate the dynamical predictions of the
 effective Maxwell equations obtained  in ~\cite{cabo} in connection
with the existence of surface waves in planar samples.
Defining experimental criteria for the detection of these waves or other
dynamical responses for the 2DEG could be helpful in determining  the
precise  effective electromagnetic equations for QHE samples.

The plan of the paper is as follows. In Section 2 the set of Maxwell
equations given  in ~\cite{cabo} are reviewed and a simple argument
is given showing that the CS action is all that is needed
for a lagrangean description of a medium having charge
conserving Hall currents.

  Section 3 is devoted to solving the equations for surface waves. The
dispersion relations and wave polarizations are explicitly found.
Only one of the solutions corresponds to a proper surface wave. The 
dispersion relation for the range of wavevectors and frequencies
in which the equations were obtained is almost linear, with a
propagation velocity coinciding with the light velocity in the 
surrounding dielectric medium. For wavevectors and frequencies 
in the above mentioned region, that is   $q\,r_o<1$, $w<w_c$
($r_o,w_c$ being the
magnetic length and the cyclotronic frequency respectively), the standard 
size of present experimental samples ($\sim 1\,mm$) is very much greater than
 the infrared optical wavelength associated with the cyclotronic frequency
at the usual values of the magnetic field. Therefore, it seems possible that
the infinite plane approximation considered here can be
satisfied in real experimental situations.
Another kind of solution also follows. It can be interpreted as
corresponding to forced oscillations of the 2DEG due charge
and current densities in the near regions.
The dispersion relation in this case shows
a gap for the spatial wavevector q below which  
real solutions for the frequency do not exist.
For low frequencies, the wavevector turns out to be almost independent
of frequency and varies only with the magnetic field. For typical
values of the field, the associated wavelength is of the order
 of $0.1\, \mu m$.

These properties of the forced oscillations motivated us to
use these solution to construct a model for  describing
special resonances recently found in experiments ~\cite{ashoo}.  In Section 4
the model is discussed. In these experiments ~\cite{ashoo}
   QHE samples were prepaired  having gate surfaces
that covered a  portion of the sample as well as a section
of the edge. After imposing a positive  gate voltage,
a metallic region (which they called the "Puddle")
 was created inside the bulk area. Results
 show the appearance of well defined and equally spaced conductance
  resonances when the  capacitance is measured
  as a function of increasing  gate voltage. It was estimated that
 the Puddle region during these resonances was separated
 from the edge zone by an incompressible strip (IS) in the QHE regime
 having a width of the order of $0.1\,\mu m$.
 This quantity is of the same order
 as the wavelength of the above discussed forced oscillations.
  The varying gate voltage can also be  imagined
   as  controlling the width of the
 incompressible strip. In addition, the capacitance was measured
 at nonzero frequency by applying an AC voltage to the gate.
 The  proposed model assumes that these resonances are related
 to the  generation of  forced oscillations by
 the AC electric fields at  values of the gate voltage
  where the width of the IS equals an integer number of halfs of
  the  characteristic
 wavelength. After selecting  fitting parameters chosen for
 a  specific value of the magnetic field, the dependence of the resonance
 maxima on the magnetic field value matches
  qualitatively the experimental data.
 A closer investigation of the consistence of this model will
 be discussed elsewhere.

 \section{ The equations: the  CS action as
  implied by Hall conductivity \& charge conservation}

  Maxwell's equations for the 4-vector potential $a_{\mu}$
describing small time dependent  electromagnetic
perturbations of the constant magnetic field $B$ in a
planar sample for the QHE regime were obtained
previously ~\cite{cabo}.
For our problem , these equations take the formm,
\begin{equation}
\begin{array}{c}
 (\nabla^2+\epsilon_d \partial^2_4) \epsilon_{\mu \nu} \,a_{\nu}
+ \nonumber \\
i \delta(x_3) \sigma \epsilon^{\alpha \mu \sigma \nu} n_{\alpha}
\partial_{\sigma}\, a_{\nu} +  \nonumber  \\
\delta(x_3) \epsilon \left( P_{\mu \nu} (u.\partial)^2-u.\partial (u_{\mu}
P_{\nu \alpha}+u_{\nu} P_{\mu \alpha}) \partial_{\alpha}+
u_{\mu} u_{\nu} P_{\alpha \beta} \partial_{\alpha}
\partial_{\beta}\right) \, a_{\nu}=0,  \label{eq:1}
\end{array}   \\
\end{equation}
\\
\noindent
where $ x_{\mu} \equiv (\vec{x},x_4),
 x_4= i x_o=i c\,t$,
$\epsilon_{\mu\nu}$ is a kind of dielectric tensor,
$P_{\mu \nu}$  is a projection operator tensor on the 2DEG
plane, $ u_{\mu}$  is the 4-velocity of
the same plane and  $n_{\mu}$,
 is a unit 4-vector being normal to the plane.
 These quantities take the
explicit forms,

$$ \begin{array}{cc}
\epsilon_{\mu \nu}\equiv \left(\matrix {
       1     &     0    &   0    &   0 \cr
       0     &     1    &   0    &   0 \cr
       0     &     0    &   1    &   0 \cr
       0     &     0    &   0    &  \epsilon_d \cr }\right),  &

P_{\mu \nu}\equiv \left(\matrix {
       1     &     0    &   0    &   0 \cr
       0     &     1    &   0    &   0 \cr
       0     &     0    &   0    &   0 \cr
       0     &     0    &   0    &   0 \cr }\right), \\

 u_{\mu}\equiv \left( \matrix{ 0 \cr 0 \cr 0 \cr 1 \cr}\right),
& n_{\mu}\equiv \left( \matrix{ 0 \cr 0 \cr 1 \cr 0 \cr}\right),
\label{eq:2}
\end{array}
$$

\noindent
where $\epsilon_d$ is the dielectric constant of the medium
where the 2DEG is embedded. For usual experimental
situations, this quantity is a number of the order of ten.
In reference ~\cite{cabo}, these equations
were obtained for the vacuum case, that is $\epsilon_d=1$. However,
here we are trying to describe  more realistic situations
that appear in mesoscopic systems. This
requires  inclusion of the dielectric properties of the
surrounding medium for the 2DEG.
These equations 
correspond to the following generalization of the Lorentz
gauge for the
4-vector potential
 \begin{equation}
  \partial_{\mu} \epsilon_{\mu \nu} a_{\nu}=
                        \partial_i a_i +\epsilon_d
                                \partial_4 a_4 =0,
\end{equation}

\noindent
which as  usual  does not fully fix.
This  allows us to impose in addition
the axial gauge
 \begin{eqnarray}
n_{\mu}a_{\mu}=a_3=0.
\end{eqnarray}

Equations (1) were obtained by performing a one loop calculation
of the dielectric tensor of the 2DEG placed in a magnetic field
$B$ for densities corresponding to the complete filling of
$l+1$ Landau levels. Only a local approximation was considered
in which  long wavelength $ r_o |\vec{k}|\ll 1$ and low
frequency $ w \ll w_c$ conditions were assumed.

 The terms in the second line of (1) represent  Chern-Simons
 sources which are equivalent to the Hall conductivity current
 and density of the system. The third line of (1) is the 4-current
 associated with the dielectric properties of the electron gas  in
 the presence of a magnetic field.
 The associated  parameters $\sigma$ and $\epsilon$ are given by
\begin{eqnarray}
\sigma&=&{4 (l+1) e^2 \over \hbar c} \\
 \epsilon&=&{ 4 (l+1) e m c
\over B \hbar}
\end{eqnarray}

 Note that the dielectric
 polarizability $\epsilon$ is inversely proportional to the magnetic field.
In (5), the mass parameter is assumed to  take the  typical
value for the effective mass for electrons in the AsGa samples equal
to $\sim 0.07$ of the electron mass. As  defined above,
 B is the external magnetic field which is of the order
 of $10$ Tesla in current
experimental conditions.

The static limit of equation (1)
reproduce many of the features of the discussion of ~\cite{mac} and
\cite{thou}, but in addition includes a
 complementary magnetic field dependent
charge density. This term was discussed for the first time
\cite{riess} as arising from the requirement of charge conservation.
In the context of our previous work \cite{cabo}, botth  approaches
are contained within equations based on a Chern-Simons
action that generates  Hall currents as well as  magnetic field
dependent charges, and where the electromagnetic fields
are considered  in 3D space.

 The set of equations (1) can be translated into a more physically
 appealing representation by expressing them in terms of the
 electric and magnetic field intensities
\begin{eqnarray}
   E_i&=& i (\partial_i a_4-\partial_4 a_i) \nonumber \\
   B_i&=&\epsilon^{i j k}
 \partial_j a_k
 \end{eqnarray}

Using  these relations and the  gauge
conditions (2) and (3) for to eliminate the $a_{\mu}$ components
from the Maxwell equations (1), the equations  can be written in the
following form

\begin{eqnarray}
\vec{\nabla}.\,(\epsilon_d \vec{E}+\delta(x_3) \epsilon
\vec{\vec{P}}.\,\vec{E}) &=& \delta(x_3) \sigma \vec{n} .\,\vec{B}
\\
\vec{\nabla} \times \vec{B} &=& \delta(x_3) \sigma \vec{E} \times
\vec{n}+{1\over c}{\partial \over \partial t}(\epsilon_d \vec{E}
+\delta(x_3) \epsilon \vec{\vec{P}}.\, \vec{E})\\
\vec{\nabla} \times \vec{E} +{1\over c} {\partial \over \partial t}
  \vec{B} &=& 0 \\
\vec{\nabla} .\, \vec{B} &=& 0
\end{eqnarray}

\noindent
where $ \vec{\vec{P}}$ is the projector on the 2DEG diadic 3D-tensor and
$\vec{n}$ is a unit vector normal to the plane.

\noindent
Equations (9) and (10)  are analogues of the vacuum  equations for
Faraday induction and the absence of magnetic charge.
In addition in eqs. (7) and (8), the $\epsilon$ dependent terms show
that the electron gas behaves as a  dielectric
surface  which  polarizes itself linearly only due to the
tangential components of the electric field. The Hall currents appear
now explicitly in the r.h.s. of (8).  The only unusual contribution is
the magnetic field dependent charge density appearing in the r.h.s. of  (7).
However, as mentioned above, an discussed in ~\cite{riess}
this term arises from the general condition of charge conservation.
Therefore, its presence  should not be
considered as an unusual outcome. Rather, such charge
densities are determined by  assuming the existence of a Hall
conductivity in a planar medium. This fact will be discussed below.

  In the set of equations (7)-(10), the  magnetic
field charge density is substituted by a undefined function $ \rho^{Hall} $.
All the other terms remain unaltered.
After taking the divergence of equation (8), and adding the result
to equation (7)  the following conservation conditions follow for
the polarization and for the Hall currents and densities
\begin{eqnarray}
{\partial \over \partial t} (\rho^{Hall}+\rho^{Pol})
  +\vec{\nabla}.(\vec{J}^{Hall}+ \vec{J}^{Pol})&=&0, \nonumber \\
{\partial \over \partial t} \rho^{Hall}
  +\vec{\nabla}.\vec{J}^{Hall}&=&0.
\end{eqnarray}

After substituting in (11) the expression for the Hall currents,
it follows that
\begin{eqnarray}
{\partial \over \partial t} \rho^{Hall} &=&  -\delta(x_3)
\vec{\nabla}.(\vec{E}\times\vec{n}) \nonumber \\
      &=&- {c\over 4 \pi} \sigma \vec{n}.\,(\vec{\nabla}\times\vec{E})
\nonumber \\
      &=&  {\partial \over \partial t}
          ({c\over 4 \pi} \sigma \vec{n}.\,\vec{B}).
 \end{eqnarray}

 This relation shows that the  unknown quantity $\rho^{Hall}$ should
differ by a time independent function from the magnetic field
dependent charge density that appears in eq. (7). Then, after
assuming that before any perturbations (by,  for example,
incoming waves) the charge
densities vanished,  it follows that the magnetic field dependent
surface charge densities (and then the whole Chern-Simons structure
of the Hall  current 4-vector) is  implied  by  the existence
of a Hall conductivity.

These are the  basic equations to be used to investigate the spectrum of
surface waves and forced oscillations for
the 2DEG in the integral  QHE regime.

\section{Surface waves and resonances }
We look for solutions of the Maxwell equations
(1) that have harmonic behavior corresponding to the
propagation of energy along the plane. To be specific, we
consider the propagation along the direction
$x_2$ of the coordinate axes. Oscillatory solutions
showing an exponential decay away from the $x_3$ axis
will correspond to  surface waves
and ones that grow outside the plane should
describe forced oscillations driven by currents and charges
at some boundary.

The vector potential will take the form

\begin{equation}
a_{\mu}(x)=a_{\mu}(k) \exp[i (q x_2 + k_3 |x_3| -
\sqrt{\epsilon_d} w \, t)],
\end{equation}

\noindent
where the wavevector component along the $x_2$ axis
will be designed as                                         
$$
k_2=q,
$$

\noindent
and all the wavevector components $q,k_3$ and $k_4=i k_o=i w/c$
are chosen to satisfy the wave equation
in the surrounding dielectric medium. That is, they obey    
\begin{equation}
q^2+k_3^2-\epsilon_dw^2/c^2=0.
\end{equation}

Relation (14) allows us to write an expression for the
wavevector component $k_3$,
which controls the behavior of the fields outside the plane,
\begin{eqnarray}
k_3&=&(-q^2+\epsilon_d w^2/c^2)^{1/2} \nonumber \\
   &=&f\, i\, (q^2-\epsilon_d w^2/c^2)^{1/2},
\end{eqnarray}

\noindent
where the positive value for the  squared root is taken when
its argument is positive. Then, the factor f  can have
two values
$$
f=\pm 1.
$$

For $f=1$ or $f=-1$, the wave amplitude decreases or grows when
away from  the surface.

Determination of the solution is simplified by considering
 the first term in the Maxwell equation (1) after substituting
the expression (13) for $a_{\mu}$. As $a_{\mu}$ satisfies the wave equation
in a dielectric medium, the result reduces to a surface contribution
on the following form
\begin{eqnarray}
(\vec{\nabla}^2+\epsilon_d \partial_4^2) \epsilon_{\mu \nu} a_{\nu}
&=&2\,i\,k_3 \delta(x_3) \epsilon_{\mu \nu} a_{\nu}.
\end{eqnarray}

Therefore, the set (1) becomes  a matrix
equation over the 2DEG plane for the
 nonvanishing  polarization components
$a_{\mu}, \mu=1,2,4$. It can be written in the form
\begin{eqnarray}
 \begin{array}{cc}
\left(\matrix {
-2 f\,X+\epsilon k_o^2  & i\,\sigma k_o & -\sigma q \cr
- i\,\sigma k_o &  -2 f\,X+\epsilon k_o^2 &   i \epsilon k_o q  \cr
  \sigma q     & i  k_o q &
-2 f\,\epsilon_d X -\epsilon q^2   \cr  }\right) &
  \left( \matrix{  a_1\cr a_2\cr a_4\cr }\right)=0,
\end{array}
\end{eqnarray}

\noindent
where $ X $ is
\begin{eqnarray}
X=(q^2-\epsilon_d w^2/c^2)^{1/2}.
\end{eqnarray}

After imposing the  condition that the
determinant of the matrix should vanish,
the following dispersion relations arise
\begin{eqnarray}
 f \sqrt{(\epsilon q)^2-\epsilon_d (\epsilon k_o)^2}&=&
       -(\epsilon_d-(\epsilon k_o)^2/4+\sigma^2/4)\nonumber \\
                    & &  \pm \sqrt{(\epsilon_d-(\epsilon k_o)^2+\sigma^2)^2
                        +\epsilon_d (\epsilon k_o)^2}.
\end{eqnarray}

From these, it can be seen that the cases  $f=1$ and $f=-1$  correspond
to the positive and negative  values  of the squared root respectively.
Then, two possibilities appear:  surface wave propagation
given by f=1 and  forced oscillations given by f=-1.

The vector potential polarization vectors can be written
 in the following way

\begin{equation}
 \begin{array}{c}
a_{\mu}= \left(\matrix {
{1\over i \sigma } [-2 f\,(q^2-\epsilon_d k_o^2)^{1/2}+
                \epsilon k_o^2-\epsilon q^2/\epsilon_d]  \cr
   k_o    \cr
   0    \cr
  - q/{ (i\,\epsilon_d)} \cr  }\right).   \label{eq: 21}
\end{array}
\end{equation}

Below, physical conditions for
the  validity of these solutions will
be discussed in each of the cases: the surface waves
and the forced oscillations

\subsection{Surface waves}

The set of Maxwell equations (1) was obtained  using the
local  approximation where the wavevector of the
perturbations $a_{\mu}$ was  assumed to satisfy
$ r_o\, q \ll 1 $ and
${ w \over c} \ll w_c$. The
 magnetic radius and the cyclotronic frequency
have the usual expressions
$ r_o = \sqrt{{\hbar c\over e B}}$ and
$ w_c = {eB \over mc}.$

Therefore, eventhough  the dispersion relation
is nonlinear, in the region of validity for the
approximations, the
 dependence  is nearly linear in  behavior.
The velocity of the waves in this low momentum zone
 coincides
with the light velocity in the surrounding dielectric medium.
 The general form of the dispersion curve is shown in Fig. 1 by the branch
crossing the origin. The whole curve gives the surface wave dispersion
for a classical  problem characterized by  planar Hall
 conductivity and dielectric responses. However,  application
  to the case of a realistic 2DEG in the QHE regime
 reduces to a narrow region where $\epsilon k_o\ll
 0.029$ for $B=6\,T$ and $\epsilon_d=10$.

Let us assume that the wavevectors obey these conditions,
and discuss the connection with the experimental
situation.

 The limitations
 imposed by $r_o q\ll 1$ are not very strong because the normal
 values for $r_o$ in present experiments are of the order
 of $100 {\cal A}^o$. Thus, up to optical wavelengths this
 approximation is well satisfied.

The more restrictive bound comes from the frequency relation
$w\ll w_c$. In this case, after  considering typical values for the magnetic
field foundin experiments $B\cong 6\, T$, the cyclotronic frequency
is  of the order of $ w_c=1.5\, 10^{13}
 $. Hence, the dispersion  relations
are expected to hold  up to the infrared region.

In connection with the possibilities for detection
of such waves under  present laboratory conditions,
it should be noted that the above  bounds on
 the frequency  are related  to
 wavelengths   $\lambda\cong 0.0039 \, cm
  $ for $\epsilon_d=10$. These values
  are  much smaller than  the normal sample sizes which can
 presently be prepared. Therefore, in this sense, the present
  analysis favors the possibility of the
  generation and detection of such waves. Clearly,
   the stability of the present picture under
  boundary effects  present in
  real mesoscopic samples should be considered in more detail.

\subsection{Forced oscillations}

The case f=-1 corresponds to oscillations which grow in the
 $x_3$ directions away from the 2DEG plane.  Let us
 consider two planar  surfaces  parallel to the $x_3$ plane
  and at the same distances, say $h$ from
  it.  Consider also, choosing the
   solutions outside these planes to be  equal
  to the corresponding  ones inside the planes  but substituting
  $k_3\rightarrow -k_3$. The special surface current and charge density
  distributions on these planes that allow such a global solutions,
   can be calculated by applying the wave equation.
    This picture allows to understand the forced nature of the
  oscillations. It also suggests that neighboring structures
   in mesoscopic systems can act as generators which force this resonant
   behavior.

  The dispersion curve associated with forced
  oscillations is ilustrated in Fig. 1 by the curve not
   intersecting the origin of the coordinate axes .
  The result shows the existence of a critical value for the
  spatial momentum $q$ below which real solutions
  for frequency are absent.
   In fact, as in the surface wave case, for the 2DEG experimental
    samples, the
  curve are  only valid for  frequencies smaller than
  $w_c$. In this region, the wavelength of the waves are almost frequency
  independent. In terms of the dimensionless
  units used in Fig.1, this region corresponds to $\epsilon k_o\ll 0.029$.

The critical wavelength for the forced oscillations is
inversely depending on the magnetic field $B$.  Its value for
normal experimental conditions $ B= 6\,T, \epsilon_d=10,$ and
$ m=0.07 m_e $ is as small as $\lambda=0.182 \mu\,m$. The
properties for such solutions ledd us to suspect taht  they are
relevant for describing  recent experiments
(\cite{ashoo}) in which special tunneling resonances were reported.

In the next section,  a model for the explanation of these
 resonances based on these forced oscillations
 will be proposed.

\section{The model for conductance  resonances}

In reference ~\cite{ashoo},  the authors  reported  experiments done
on QHE samples in which a gate surface
covered  part of the bulk  as well as part
of the edge. After applying a positive  gate voltage,
a metallic region (the "Puddle")
 was created inside the bulk . It was composed of
electrons excited to the partially filled next Landau
level by the attracting gate voltage. The capacitance between
the gate and the edge contacts was measured as a function
 of the applied gate voltage for a range  of values of the
 frequency and the  magnetic field. The results
 show the appearance of equally spaced resonances
 when the  gate voltage was increased. It has been estimated that
 the Puddle region  should be separated
 from the edge  by an incompressible strip (IS) in the QHE regime
 having a width of the order of $0.1 \mu m$. A reasonable
 phenomenological model used by the authors indicates that
 these resonances are associated with  resonant
 tunneling currents passing across the IS.

Below, we give arguments suggesting
that the  forced oscillations discussed in last section
are candidates for producing  these resonances.

The reasoning goes as follows. As noticed before, the low
frequency spectrum of forced oscillations is characterized by
a constant wavelength which is given by
\begin{eqnarray}
\lambda &=& {\pi \epsilon \over \epsilon_d +\sigma^2/4} \nonumber \\
        &=& { 4 \pi (l+1) e m c \over \hbar B (\epsilon_d+
          \sigma^2/4)}.
\end{eqnarray}

In experiments the IS is
subject to a time varying voltage which is applied to
measure the capacitance. The spatially inhomogeneous form of this
potential can be considered as the source for the external charges which
necessary for the  forced oscillations to be excited.

We consider now the following  assumptions:

1). The IS
only resonates when its length L coincides with an integer number
of half wavelength of  the forced wave.

2) The odd symmetry of the applied potential
with respect to the center of the plate restricts the
the resonating condition requiring
the length L to correspond to an odd number of halfs wavelength.

3) Increasing the gate voltages reduces the width L of the
IS, with the various resonances being produced
according to the previous rules.

The main elements elements considered within the model are ilustrated
in Fig. 2.

Under the above suppositions, we can write
the condition which links the length of the strip with the magnetic
field as follows
\begin{eqnarray}
{2 \pi L \over \lambda}=(2 n+1) \pi.
\end{eqnarray}

However, the experimental data  ~\cite{ashoo} are given in terms of
the dependence of the resonances on the gate voltage $U_g$. Thus,
in order to compare the consequences of the model with experiment,
it is convenient to estimate the functional link between the
 length L of the IS
and  the gate voltage  $U_g$, by fitting some experimental data.
For this purpose, we assume a linear relationship between $ L $ and
$U_g$  as
\begin{eqnarray}
L=\alpha (U_o-U_g),
\end{eqnarray}

\noindent
where $ U_o$ will also be fixed by estimating the voltage $U_g$
at which the Puddle joints with the edge.

Taking the voltage shift between resonances as equal to $0.040 \, mV$
 ~\cite{ashoo}, the parameter $\alpha$ is given by

\begin{eqnarray}
\alpha={\lambda\over 0.040}.
\end{eqnarray}

Restricting  to the first Landau level $l=0$, the voltage
$U_o$ can be evaluated from Fig. 2  in Ref. ~\cite{ashoo} to be
approximately $ U_o\cong 350\, mV$

Next, we defined  a function of the variable
 $ x=L -(n+1/2)\lambda $ to be a
 sum of six gaussian functions having maxima at each integral value of $
x=0,1,..5$. The variable $x$ is expressed
 in terms of the magnetic field $B$ and the gate voltage
$U_g$ by using (22) and (23) in order
to investigate the behavior of this
function of  $U_g$ as
the magnetic field is varied. The result is shown in Fig. 3 and
reproduces qualitatively the data of Fig. 2 in ~\cite{ashoo}.
The most interesting fact is that
the slopes for the dependence of the maxima on the magnetic field match.
 The fitting parameters taken from the experiment  were the voltage
 separation between resonances at the value of $B= 6\, T$ and the voltage
 $U_o\cong 350\, mV$ for the disappearance  of the incompressible strip
 at the same field value.
 The  description of the slopes is a  prediction of the
 model which introduces a dependence of the wavelength on the magnetic
 field  that qualitatively agrees with  the experiment.

It can also be added that the proposed approach naturally explains the
 global character of the resonances which appear for various
sizes of the gate electrode. In future work, a derivation of the
main assumptions adopted above will be studied.

\section{Acknowledgments.}

One of the authors (A.C.) greatly acknowledge the support of the
High Energy Section of the ICTP for a visit which was central in
allowing to perform this work and  to link it with  recent experimental
activity on the subject and also greatly appreciate the invitation of the 
Condensed Matter Section to participate in the very helpful Adriatico 
Research Conference on: "The Electron Quantum Liquid in Systems of Reduced
Dimensions"(2-5 July 1996) .  The remarks and help of E. Tosatti, J. Lawson, G. 
Ernst and A.H. MacDonald are also very much acknowledged.

We all are also deeply indebted  to the Third World Academy of
Sciences for its support through  the TWAS Research Grant
93 120 RG PHYS LA.
\newpage


\newpage

{\large \bf Figure Captions}

\begin{itemize}
\item{\bf Fig. 1}\, \\
         The surface wave (starting at the origin) and forced oscillations
         dispersion relations plotted in terms of the dimensionless wavevector
         components $\epsilon q$ and $\epsilon k_o$. The parameters chosen were:
         $B=6\,T,\epsilon_d=10, m=0.07 m_e$. Only in a small frequency region
         $\epsilon k_o<0.029$ do the dispersion relations
         satisfy the conditions imposed by the local
         approximation for the dielectric response.

\item{\bf Fig. 2}\,\\
         A picture of the main elements of the
         model for conductance resonances.

\item{\bf Fig.3}\, \\
         The graphical representation for the dependence on the
         gate  voltage and the external magnetic field of the functions
         representing the resonance maxima predicted by the model.
         The fitting parameters were the voltage difference between
         the maxima at $B=6\,T$ and the voltage $U_o$ at which the
         incompressible strip seems to disappear at the same magnetic field
         value of $6\, T$.
\end{itemize}

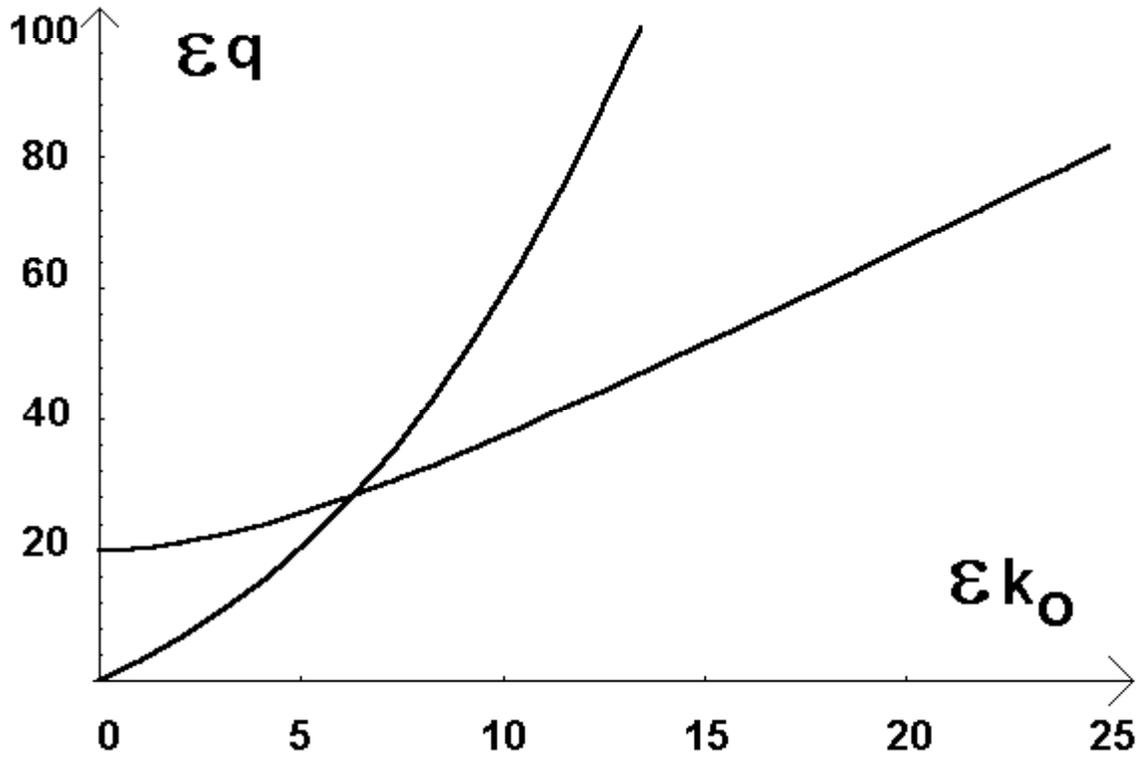
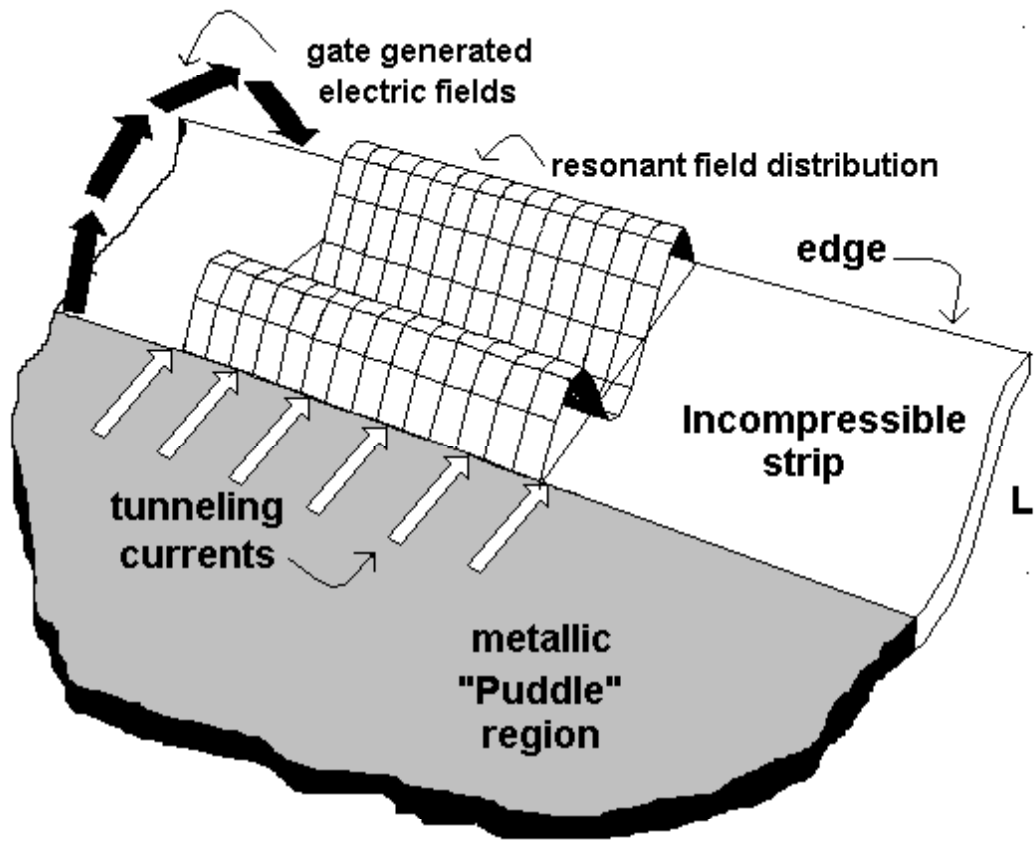
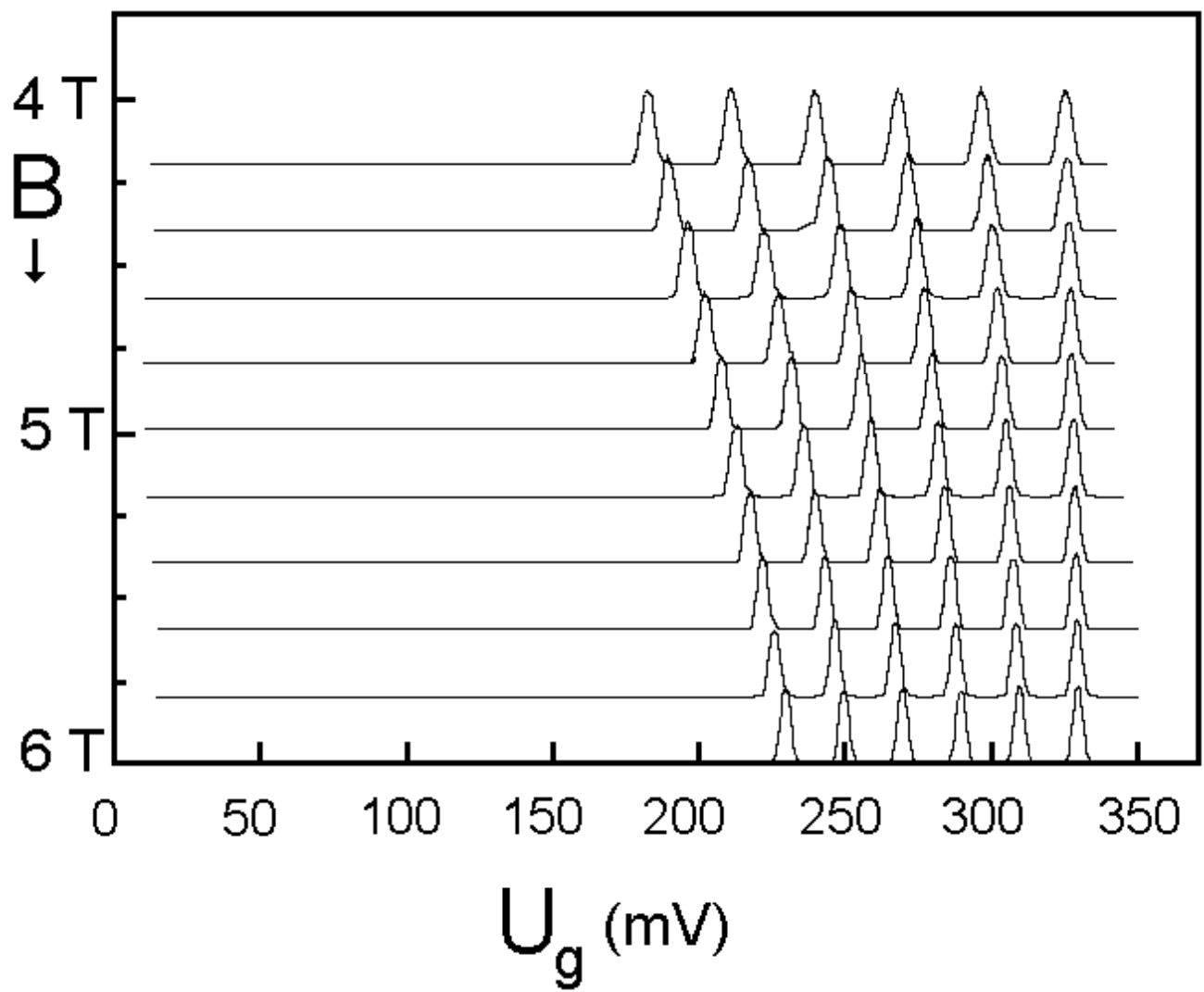
\end{document}